\documentclass[aps,twocolumn,showpacs,pra,superscriptaddress]{revtex4}
\usepackage{graphicx}
\usepackage{amsfonts}
\newcommand{\ra}{\rangle}
\newcommand{\la}{\langle}
\newcommand{\tbf}{\textbf}

\begin{document}

\title{Evaporative cooling in a radio--frequency trap}
\author{Carlos L. Garrido Alzar}
\email{leonardo@galilee.univ-paris13.fr}
\affiliation{Laboratoire de Physique des Lasers, CNRS-Universit\'e
  Paris 13, Villetaneuse, France}
\author{H\'el\`ene Perrin}
\affiliation{Laboratoire de Physique des Lasers, CNRS-Universit\'e
  Paris 13, Villetaneuse, France}
\author{Barry M. Garraway}
\affiliation{Department of Physics and Astronomy, University of Sussex,
  Brighton BN1 9QH, United Kingdom}
\author{Vincent Lorent}
\affiliation{Laboratoire de Physique des Lasers, CNRS-Universit\'e
  Paris 13, Villetaneuse, France}
\date{\today}

\begin{abstract}
 A theoretical investigation for implementing a scheme of forced evaporative cooling in
 radio--frequency (rf) adiabatic potentials is presented. Supposing the atoms
 to be trapped by a rf field at frequency $\omega_1$, the cooling
 procedure is facilitated using a second rf source at frequency $\omega_2$. This
 second rf field produces a controlled coupling between the spin
 states dressed by $\omega_1$. The evaporation is then possible in a
 pulsed or continuous mode. In the pulsed case, atoms with a given
 energy are transferred into untrapped dressed states by abruptly
 switching off the interaction. In the continuous case, it is possible
 for energetic atoms to adiabatically follow the doubly--dressed states
 and escape out of the trap. Our results also show that when $\omega_1$ and $\omega_2$
 are separated by at least the Rabi frequency associated with $\omega_1$,
 additional evaporation zones appear which can make
 this process more efficient. 
\end{abstract}

\maketitle

\section{Introduction}
In recent years, the investigation of quantum gases in low--dimensional trapping
geometries has significantly attracted the attention of the physics
research community. This growing interest is motivated,
partially, by the current possibilities that the extremely rapid progress in
integrated atom optics has opened for the manipulation of
Bose--Einstein condensed (BEC) atoms. This development allows the study of
crucial problems associated with the strong modifications that the fundamental
properties of these quantum systems experience due to the reduced
dimensionality. For instance, a 1D Bose gas in the Tonks--Girardeau
regime mimics a system of non--interacting spinless
fermions~\cite{tonks,girardeau,paredes}; in 2D, the superfluidity emerges
due to the vortex binding--unbinding Berezinskii--Kosterlitz--Thouless
phase transition~\cite{BKT1,BKT2}, recently observed~\cite{zoran}.

For the study of the BEC low--dimensional physics, trapping
configurations of different nature and topology have been proposed and
used. For example, the 3D to 1D crossover was explored by G\"orlitz \emph{et
  al}.~\cite{gorlitz} in an elongated
Ioffe--Pritchard type direct--current (dc)
magnetic trap, the phase defects of a
BEC were investigated in a quasi--2D trap based on a 1D optical
lattice~\cite{stock} and, in atom chip experiments, dc
current--carrying wires are usually employed to confine
atoms in highly anisotropic traps~\cite{folman}. Although these trapping configurations have
demonstrated their relevance for studying quantum gases in low
dimensions, adiabatic potentials~\cite{zobay},
resulting from a combination of dc and radio--frequency (rf) magnetic
fields, are also becoming a very attractive and promising
tool~\cite{colombe,schumm,morizot,courteille,lesanovsky,fernholz}. 

The rf traps share the versatility and flexibility of the above
mentioned trapping schemes and, moreover, they are relatively easy to
implement and control. In the first implementation of these
traps~\cite{colombe}, ultra--cold atoms were confined in a 2D
geometry. A rf adiabatic potential has also been used as a beam
splitter, allowing the demonstration of matter--wave interference on
an atom chip~\cite{schumm}. Ring--shaped traps, and other more complex
trapping geometries using adiabatic potentials have also been
considered~\cite{morizot,courteille,lesanovsky,fernholz}.

Given the topology of the rf trapping potential, and because of
technical limitations in some cases, the loading of the trap with
Bose--Einstein condensed atoms, preserving the quantum degeneracy, can
be a challenging task. In this situation, it is of relevance to
consider the possibility of evaporative cooling of atoms \emph{directly}
in these low--dimensional rf traps. This is the subject that will be
addressed in this paper, taking into account the interaction of the
atoms with two radio--frequency fields. When dealing with more than one rf
frequency, an analytical solution for the atomic spin dynamics can be found
by treating the individual successive interactions of the rf fields with the
atoms~\cite{ramsey} or by considering the two fields simultaneously,
provided one of the fields is rather weak~\cite{ficek}. We will study how a weak second
radio--frequency source can be used to perform an evaporation. 

We will see that the forced evaporation of rf--trapped atoms can be
accomplished in two ways. Firstly, the spin
evolution induced by this second rf source can be quenched by
switching off the field, \emph{i.e.} by using a
pulsed rf source. Secondly, we can allow an adiabatic following of
doubly--dressed states which requires the second rf source to be
continuous rather than pulsed. This last scheme is similar to the
standard evaporative cooling method used in static magnetic
traps~\cite{hess,vitanov,luiten}. 

This paper is organized as follows: In section~\ref{sec:conf} we will
discuss the geometry of the system and the singly--dressed states of
the rf trap. In Sec.~\ref{sec:dyna}, the evolution of the system is determined
in three different ways: numerically, using a first order Magnus
series approximation, and by using a second rotating wave
approximation which leads to a double--dressing of the
atoms by two rf fields. Sec.~\ref{sec:evap} is devoted to the application of the
results of Sec.~\ref{sec:dyna} to the study of evaporative cooling in
the rf trap. Finally, we give a summary and conclusion in Sec.~\ref{sec:conclu}.

\section{Adiabatic potential confinement}\label{sec:conf}
The underlying idea of the confinement of ultra--cold atoms using
rf adiabatic potentials is presented in detail in Ref.~\cite{zobay},
Ref.~\cite{colombe} being the first report on the
experimental investigation of such a trapping scheme. For this reason, instead of
discussing deeply how this trapping actually takes place, we will rather
make use of the already known results that are relevant in order to consider the
problem of evaporative cooling in these traps. 

The treatment presented in this paper is valid for any value of the
spin $F$. However, for concreteness the numerical results will be given for
$^{87}$Rb in the Zeeman state $m = +2$ of the $5S_{1/2}$, $F = 2$
hyperfine ground state level. We will suppose that the atoms are confined
in a QUIC magnetic trap~\cite{esslinger, lu} produced by a
dc magnetic field $\tbf{B}_{dc}(\tbf{r})$. The atomic clouds trapped in this configuration are
anisotropic (cigar--shaped along $x$) and we will take the offset magnetic field produced
by the Ioffe coil~\cite{esslinger} to be oriented along the $x$ direction. In the
following, the axes in the lab frame will be labelled by lower case
letters $xyz$. The $z$ axis is in the vertical direction and $y$ is the horizontal
direction perpendicular to the cigar axis. The axes in the local frame
attached to the dc magnetic field will be labelled by capital letters $XYZ$. Moreover, we will assume
that the direction of the dc magnetic field defines the local $Z$
quantization axis. The Larmor frequency of the atomic spin
precession in this dc field will be denoted $\omega_0(\tbf{r}) = g_F
\mu_B B_{dc}(\tbf{r})/\hbar$. Here, $g_F$ and $\mu_B$ are the Land\'e factor and the
Bohr magneton, respectively.

Now, we apply to this confined atomic system two rf fields (produced
by antennae), both of them polarized along $X$ and of angular frequencies
$\omega_1$ and $\omega_2$. Having in mind that the second rf field
will be rather weaker than the first one, we transform into the frame
rotating at $\omega_1$ and perform the rotating wave approximation (RWA). The
Hamiltonian that describes the spin dynamics can thus be written as (see
the derivation in appendix~\ref{app:Hamiltonian})
\begin{equation}
 H(\tbf{r},t) = H_A(\tbf{r}) + \Omega_2 \big[F_X \cos(\Delta t) + F_Y \sin(\Delta t)\big] \ ,
\label{eq:iniH}
\end{equation}
where $F_X (F_Y)$ is the atomic angular momentum in the $X (Y)$
direction, $\Delta = \omega_2-\omega_1$, $H_A(\tbf{r}) = \Omega(\tbf{r})
F_\theta$ is the adiabatic Hamiltonian associated with the rf
confinement, and \mbox{$\Omega(\tbf{r}) \equiv
  \sqrt{\delta(\tbf{r})^{2}+ \Omega_1^{2}}$} defines the energy separation between the adiabatic
levels. In the absence of $\Omega_2$, the flip angle $\theta$ and the detuning $\delta(\tbf{r})$ are
given by \mbox{$\tan(\theta) \equiv -\Omega_1/\delta(\tbf{r})$}, with
$\theta\in[0,\pi]$, and
\mbox{$\delta(\tbf{r}) = \omega_1-\omega_0(\tbf{r})$},
respectively. We have labelled the Rabi frequencies of the rf fields
$\Omega_1$ and $\Omega_2$, and we have considered that the component $F_Z$ of the atomic angular
momentum is aligned with the local $Z$ component of the dc magnetic
field vector. Strictly, these Rabi frequencies are not spatially homogeneous, however they can be
treated as such over the spatial extension of the atomic cloud~\cite{colombe}.

\begin{figure}[htb]
\begin{center}
 \includegraphics*[width=0.38\textwidth]{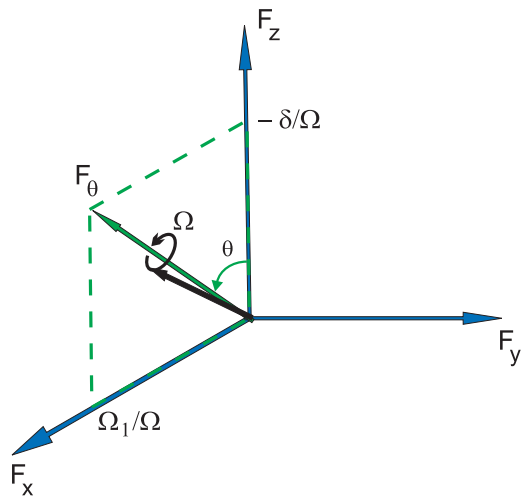}
\caption{(color online). At a given location $\tbf{r}$, $\delta$ and $\Omega_1$ define the
 angle $\theta$ by which $F_Z$ is rotated. The spin (black arrow)
 therefore precesses around an axis given by $F_\theta$ (green arrow)
 at a frequency $\Omega$.}
\label{fig:1}
\end{center}
\end{figure}

Graphically, the spin evolution given by Eq.~(\ref{eq:iniH}), in the case
$\Omega_2 = 0$, is represented in Fig.~\ref{fig:1}. It can be seen in
this figure that the tilted angular momentum $F_\theta$ results from a
rotation of $F_Z$ around $F_Y$ and is given by
\begin{equation}
 F_\theta = \cos(\theta)F_Z + \sin(\theta) F_X =
 \mathcal{R}_Y(\theta) F_Z \mathcal{R}_Y^{\dag}(\theta)\ ,
\label{eq:rotF}
\end{equation}
where the rotation matrix $\mathcal{R}_Y(\theta)=\exp(-i\theta
F_Y/\hbar)$ can be expressed in the basis $\{-2,...,+2\}$ of the
bare states.

\begin{figure}[htb]
\begin{center}
 \includegraphics*[width=0.48\textwidth]{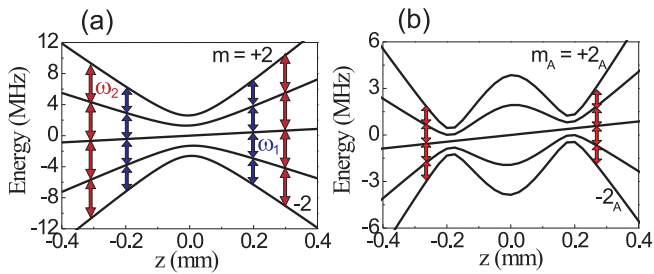}
\caption{(color online). Spatial $z$ dependence of the
  energy of the uncoupled (a) and dressed (b) states for $x = x_{min}$
  and $y = 0$. At the avoided level crossings, the energy splitting
  between the levels is $\Omega_1$. Gravity has been taken into
  account in both cases and $\omega_2 > \omega_1$.}
\label{fig:2}
\end{center}
\end{figure}

In Fig.~\ref{fig:2}(a), the energies of the bare states are plotted
as a function of the position $z$, where the energy variation is due
to the dc magnetic field $\tbf{B}_{dc}(\tbf{r})$. This spatial dependence has been calculated
for a value $x = x_{min} = 6.9$~mm corresponding to the position where the QUIC magnetic field is
minimal in our experimental setup~\cite{colombe}. Moreover, we have taken $y = 0$ and a Rabi frequency
$\Omega_1/2\pi = 400$~kHz. The arrows shown in that figure, in
blue for $\omega_1$ and in red for $\omega_2$ (each frequency stands
to the right of its respective arrows), indicate the locations
where the corresponding rf fields resonantly couple the states in
the laboratory frame. On the other hand, the spatial $z$ dependence of the
adiabatic states internal energies is shown in Fig.~\ref{fig:2}(b) where the
states are labelled, from top to bottom: $|+2_A\ra,\ |+1_A\ra,\ |0_A\ra,\ |-1_A\ra$,
and $|-2_A\ra$. We can also see in this last figure the avoided level
crossings at the positions where $\omega_1$ (taken equal to $2\pi\times$3.19 MHz
in this example and from now on) resonantly couples the bare states.

In order to consider only the confinement ($\Omega_2 = 0$) using the adiabatic potentials
shown in Fig.~\ref{fig:2}(b), let's suppose that initially we have a
$m = +2$ spin polarized ultra--cold atomic sample. In this situation, the trapping potential
corresponding to the bare state $|+2\ra$ can be adiabatically deformed
into the one associated with the dressed state $|+2_A\ra$. Such a
transformation can be performed by changing the detuning
$\delta(\tbf{r})$ from red to blue at constant Rabi frequency
$\Omega_1$~\cite{zobay,colombe} or, by increasing $\Omega_1$ at a constant red
detuning~\cite{schumm}. Here, by adiabatic deformation we mean that the
angular precession frequency $\Omega(\tbf{r})$ of the spin in
Fig.~\ref{fig:1} must be much larger than the rate at which the angle
$\theta$ changes [$|\dot{\theta}| \ll \Omega(\tbf{r})$]. Using the
loading schemes just mentioned, it is
possible to obtain highly anisotropic rf traps with trapping
frequencies, in the strongest confinement direction, ranging from
several hundreds of Hz up to a few kHz.

Having described the main properties of the adiabatic confinement,
let's now address the following issues. Assuming intuitively the existence of the
resonances represented by the arrows in Fig.~\ref{fig:2}(b), we would
like to know precisely where they are located and how strong 
they are. Another relevant point to be taken into account concerns the effect of these resonances
at the rf trap centre when, numerically, $\omega_1$ and $\omega_2$ are
close to each other. Moreover, it will be interesting to find out the different parameter
values for which the second rf field induces transitions between the
adiabatic states, in the perturbative limit with $\Omega_2 \ll \Omega_1$, leading to a possible
implementation of evaporative cooling in rf traps.

\section{Dynamics of the system}\label{sec:dyna}
In this section we will study the dynamics of the system using three
different methods. In the first case, the exact numerical
solution of the time--dependent Schr\"odinger equation (TDSE) will be
found. Secondly, an approximated analytical treatment will be
presented (Magnus approximation) in order to interprete the exact results derived
numerically. Finally, we will present an analytic solution based on a
second rotating wave approximation which will be the basis of the
analysis presented in Sec.~\ref{sec:evap}. 

Since the evolution of the atomic external and internal degrees
of freedom takes place on very different time scales, here we will
decouple the two dynamics and consider only the time evolution of the
internal degrees of freedom.

\subsection{Numerical solution}\label{sec:numer}
\begin{figure}[htb]
\begin{center}
\includegraphics*[angle=0,width=0.48\textwidth]{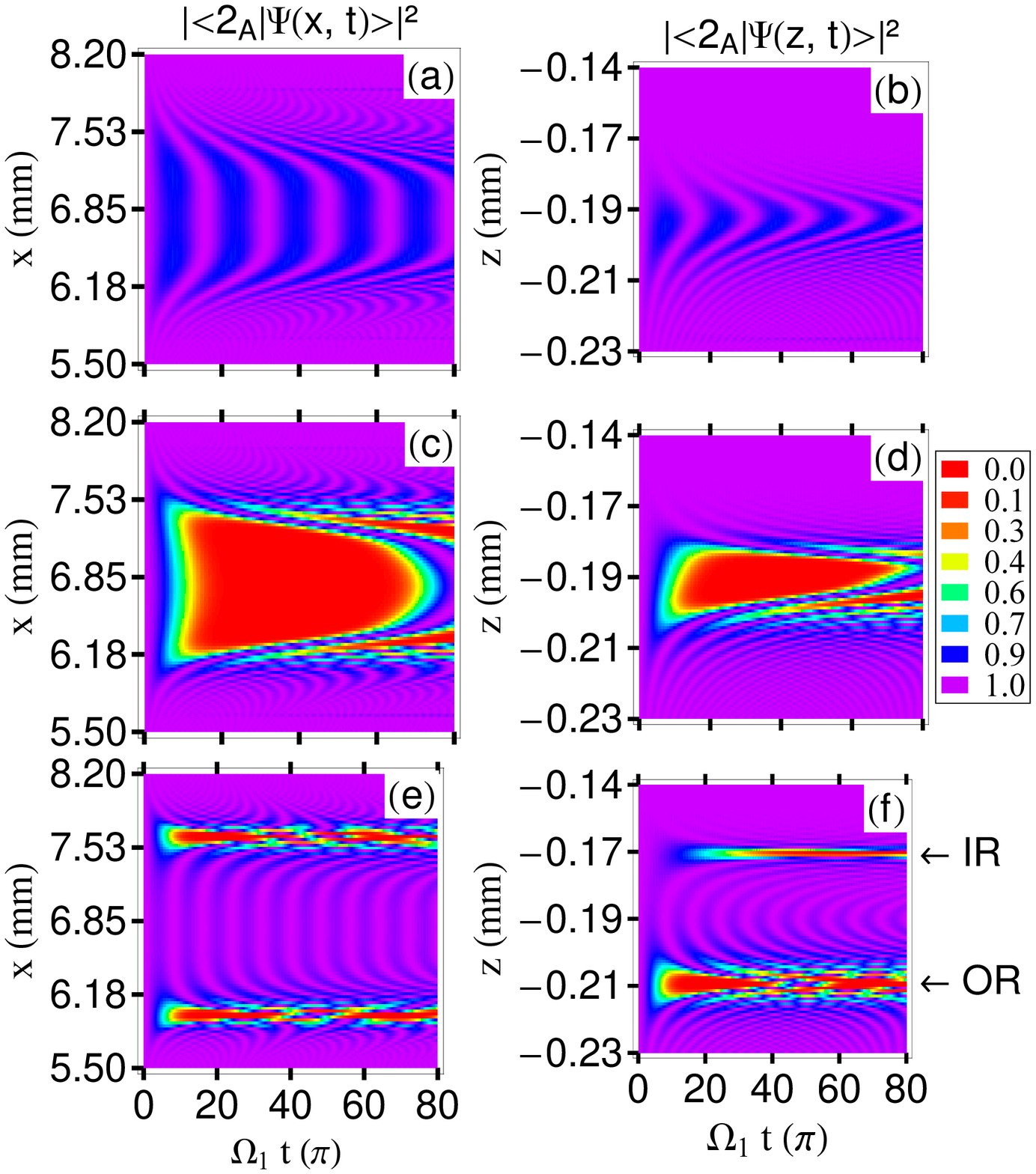}
\caption{(color online). Probability for an atom to remain in the initial rf trapped
  state $|+2_A\ra$. The values of $\Delta$ are respectively:
  0.875$\Omega_1$ in (a) and (b); $\Omega_1$ in (c) and (d); and
  1.25$\Omega_1$ in (e) and (f). The rf trap frequency $\omega_1/2\pi$
  is set to 3.19 MHz. The $x$-$t$ ($z$-$t$) dependence of $P_{2A}$ is
  calculated at $z_{min} = -0.19$ mm ($x_{min} = 6.9$ mm). In (f) the
  labels OR and IR indicate the locations of the outer and inner
  resonances, respectively.}
\label{fig:3}
\end{center}
\end{figure}

The evolution of the state vector $|\Psi(t)\ra$ with the Hamiltonian~(\ref{eq:iniH}) was
solved numerically, in the interaction picture, using a 4th order
Runge--Kutta algorithm. In this case, the
state vector can be very efficiently propagated in time and we do
not expect to have convergency problems if we choose an appropriate
time--step~\cite{tremblay}. The first question we would like to address here
is: supposing that an atom is initially in the trapped dressed state
$|+2_A\ra$, what is the probability $P_{2A} = |\la 2_A|\Psi(\tbf{r},
t)\ra|^2$ of finding it in that \emph{same} state as
time goes by? We will also be interested in how this probability
changes for an atom located in different places in the trap. The preliminary answer to these
questions is presented in Fig.~\ref{fig:3}, where the probability we are
interested in is plotted for three different values of $\Delta$, the difference 
between the radio--frequencies $\omega_1$ and $\omega_2$.

In Fig.~\ref{fig:3}, and from now on, we use $\Omega_1$ as a
frequency unit, having in mind to keep it constant in a given
experimental situation. The value of $\Omega_2$ has been taken equal
to 0.05$\Omega_1$, small enough in order to be in the perturbative
limit, as will be shown later. This value of $\Omega_2$ will be
used in all the following results unless a different one is
explicitly stated. In the left column of Fig.~\ref{fig:3}, the
$x$ and time dependence of $P_{2A}$ are calculated at the points $y =
0$ and $z_{min} \approx -0.19$ mm, with this value of $z$ corresponding to the location
of the left avoided level crossing in Fig.~\ref{fig:2}(b). To obtain
the right column of figure 3, we use for $x$ the value $x_{min}
\approx 6.9$ mm, very close to the location where the QUIC trap has its minimum in the
$x$ direction. 

In Fig.~\ref{fig:3} the three values of $\Delta$ have been chosen to
illustrate some key characteristics of the spin dynamics. As the energy
separation between the adiabatic levels at the avoided level crossings
is exactly $\Omega_1$, we observe resonant behaviour at the rf trap
bottom for $\omega_2 = \omega_1 + \Omega_1$, i.e.\ when $\Delta=\Omega_1$ 
[see Fig.~\ref{fig:3}(c) and (d)]. Similarly, for $\Delta < \Omega_1$
we have a red--detuned interaction everywhere [Fig.~\ref{fig:3}(a) and
(b)]. In this case we observe weak modulations of $P_{2A}$ that are essentially determined by a
beating between the $\Delta$ and $\Omega(\tbf{r})$ frequency
components. For $\Delta > \Omega_1$ [Fig.~\ref{fig:3}(e) and (f)] we have a
blue--detuned interaction around the minima $x_{min}$ and
$z_{min}$. Away from the centre a resonance occurs, as
expected, at approximately the location in the trap where $\omega_2$ resonantly
couples the bare states [see Fig.~\ref{fig:2}(a)]. This is the outer
resonance labelled OR in Fig.~\ref{fig:3}(f). However, there is another feature we would
like to stress. Namely, the presence of the \emph{inner resonance} IR
clearly seen in the $z$ dependence of $P_{2A}$ in
Fig.~\ref{fig:3}(f). Note that the avoided level crossing (rf trap
centre) is at $z_{min}$. Looking back to Fig.~\ref{fig:2} and
having in mind that $\omega_2 > \omega_1$, the
existence of this second resonance IR in Fig.~\ref{fig:3}(f) at $z >
z_{min}$ may be counter--intuitive and, as we
will see, its relative strength is fully determined by the rotation angle
$\theta$. Note that because of the loose
confinement in the $x$ direction~\cite{colombe}, the dynamics in the $y$-$z$ plane does
not change much from one location to another in the $x$ axis. However,
this dynamics is very sensitive to changes in $z$ (or $y$) and
therefore, the results for the $x$ dependence of the probability
$P_{2A}$ in Fig.~\ref{fig:3}(a), (c), and (e) can
be significantly different when another $z$ location is considered.

\subsection{Interpreting the numerical results using a first order
  Magnus series approximation}\label{sec:Magnus}
Searching for the understanding of the physical picture behind the numerical results
presented in Fig.~\ref{fig:3}, let's consider the first order Magnus
series approximation~\cite{magnus, milfeld} to the solution of
the TDSE. This approximation is basically the formal
solution of the TDSE neglecting the two--time commutators of the
Hamiltonian. This Hamiltonian is given in the interaction picture by
\begin{equation}
 H^{'}(t) = \exp(i \Omega t F_\theta/\hbar) H(t) \exp(-i \Omega t
 F_\theta/\hbar)\ .
\label{eq:rotH}
\end{equation}

In Eq.~(\ref{eq:rotH}), we have dropped the $\tbf{r}$ dependence in
$\Omega$ and $F_\theta$ for the
sake of notational simplification. Making use of Fig.~\ref{fig:1} and
introducing $F_{\perp \theta} = \mathcal{R}_Y(\theta) F_X
\mathcal{R}_Y^{\dag}(\theta) = \cos(\theta) F_X - \sin(\theta) F_Z$
(see Fig.~\ref{fig:6}), which is the angular momentum vector perpendicular
to $F_\theta$ and $F_Y$, we find
\begin{widetext}
\begin{eqnarray}
 H^{'}(t) = \Omega_2 \Big\{ \sin(\theta) \cos(\Delta t) F_\theta + 
 \big[\sin(\Delta t) \sin(\Omega t) + \cos(\theta) \cos(\Delta t)
 \cos(\Omega t)\big] F_{\perp \theta} + \nonumber \\
 \big[\sin(\Delta t) \cos(\Omega t)-\cos(\theta) \cos(\Delta t)
 \sin(\Omega t)\big] F_Y \Big\} \ .
\label{eq:rotH1}
\end{eqnarray}
\end{widetext}

In this case, the time evolution of a given
initial dressed spin state $|\Psi(0)\ra$ is represented by the rotation
\begin{equation}
 |\Psi(t)\ra=\exp(-i\ \mathbf{\Xi}(t)\cdot\tbf{F}/\hbar) |\Psi(0)\ra \ ,
\label{eq:magnussol}
\end{equation}
where the scalar product $\mathbf{\Xi}(t)\cdot\tbf{F}$ stands for $\Xi_\theta(t) F_\theta + \Xi_{\perp \theta}(t)
F_{\perp \theta} + \Xi_Y(t) F_Y$, where
$\Xi_\theta(t)$, $\Xi_{\perp \theta}(t)$ and $\Xi_Y(t)$ are a measure of the
projections of the spin $|\Psi(t)\ra$ precession axis onto the axes
$F_\theta$, $F_{\perp \theta}$, and $F_Y$, respectively (for
simplicity we will just call them projections). Taking into
account the equation (\ref{eq:rotH1}), the definition of the exponential argument
in Eq.~(\ref{eq:magnussol}) is
\begin{equation}
 \mathbf{\Xi}(t)\cdot\tbf{F} = \int_0^{t}dt^{'} H^{'}(t^{'}) \ ,
\label{eq:magnusint}
\end{equation}
leading to the following expressions for the projections
\begin{widetext}
\begin{equation}
 \Xi_\theta(t)=\frac{\Omega_2}{\Delta} \sin(\theta) \sin(\Delta t)\ ,
\label{eq:magnusproj1}
\end{equation}
\begin{equation}
 \Xi_{\perp \theta}(t)=\Omega_2 \Big\{\cos^{2}(\theta/2)\frac{\sin[(\Delta-\Omega)t]}{\Delta-\Omega}-
 \sin^{2}(\theta/2)\frac{\sin[(\Delta+\Omega)t]}{\Delta+\Omega}\Big\}
 \ ,
\label{eq:magnusproj2}
\end{equation}
\begin{equation}
 \Xi_Y(t)=2 \Omega_2 \Big\{\cos^{2}(\theta/2)\frac{\sin^{2}[(\Delta-\Omega)t/2]}{\Delta-\Omega}+
 \sin^{2}(\theta/2)\frac{\sin^{2}[(\Delta+\Omega)t/2]}{\Delta+\Omega}\Big\} \ .
\label{eq:magnusproj3}
\end{equation}
\end{widetext}

\begin{figure}[b]
\begin{center}
\includegraphics*[angle=-90,width=0.47\textwidth]{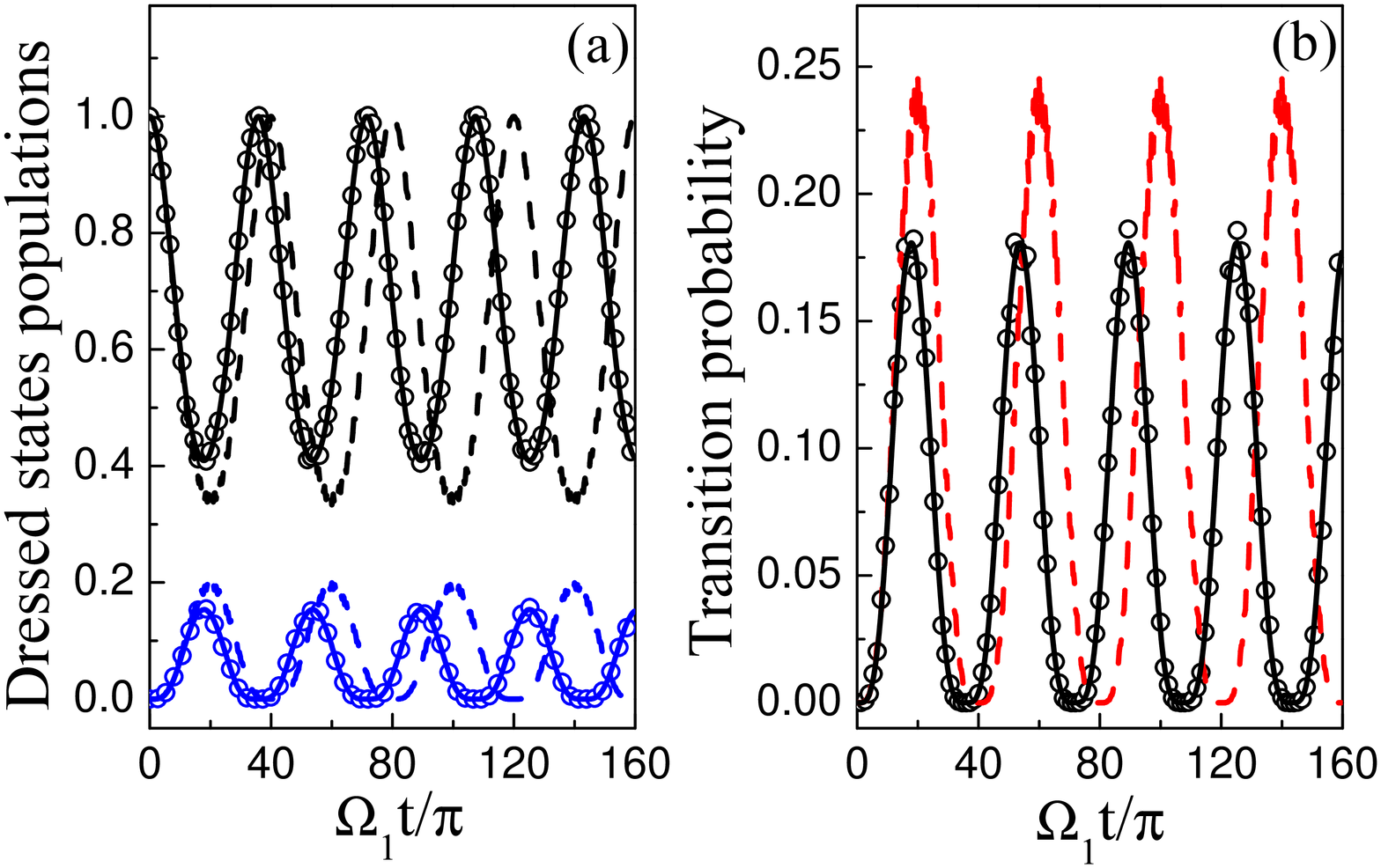}
\caption{(color online). Occupation probability of the dressed states
  $|2_A\ra$ (black upper traces) and $|0_A\ra$ (blue lower traces) (a). In (b) the probability for
  an atom to be in the untrapped dressed states is shown. The circles
  show exact numerical results  (Sec.~\ref{sec:numer}), the dashed
  lines from the Magnus approximation (Sec.~\ref{sec:Magnus}), and the
  solid lines from the second RWA (Sec.~\ref{sec:2RWA}). The
  calculation has been done at the rf trap bottom ($x_{min}, 0, z_{min}$)
  for $\Delta = 1.05\Omega_1$ and $\Omega_2 = 0.05\Omega_1$.}
\label{fig:4}
\end{center}
\end{figure}

By inspecting Eqs.~(\ref{eq:magnusproj1})--(\ref{eq:magnusproj3}), we
can see the time--dependent terms resulting from a beating between
frequency components at $\Delta$ and $\Omega(\tbf{r})$. These beats
are seen as the modulation (interference--like patterns) of $P_{2A}$
observed in Fig.~\ref{fig:3}. In particular, we see in these equations
that there will be some interesting behaviour when $\Delta = \pm
\Omega$. In either case the condition is realized by two values of $\theta$: $\theta_0$ and $\pi-\theta_0$
with $\theta_0 = \arcsin(\Omega_1/\Delta)$. If $\Delta \neq \pm\Omega$, the coefficients $\Xi_i(t)$ are
oscillatory with finite amplitudes. However, when $\Delta = \Omega$ at
$\theta = \theta_0$, for instance, $\Xi_{\perp \theta}(t)$ shows a linear tendency in time of
the form $\Omega_2 \cos^2(\theta_0/2) t$ while the other two components
are negligible. This suggests that the spin will essentially rotate
at a frequency $\Omega_2 \cos^2(\theta_0/2)$ around the axis $F_{\perp
\theta}$. Starting from an eigenstate of $F_\theta$, as in
Sec.~\ref{sec:numer}, the spin will be completely flipped after a half
period. This rotation corresponds to the outer resonance OR in
Fig.~\ref{fig:3}(f). At the location $\theta=\pi-\theta_0$, the same
resonant behaviour occurs with a rotation frequency $\Omega_2
\sin^2(\theta_0/2)$. This corresponds to the inner resonance IR in
Fig.~\ref{fig:3}(f). In the case when $\omega_2$ is smaller than
$\omega_1$, we have the resonant condition $\Delta = - \Omega$ and we
have the same behaviour except that the character of the inner and
outer resonances is now reversed.

\begin{figure}[b]
\begin{center}
\includegraphics*[angle=-90,width=0.48\textwidth]{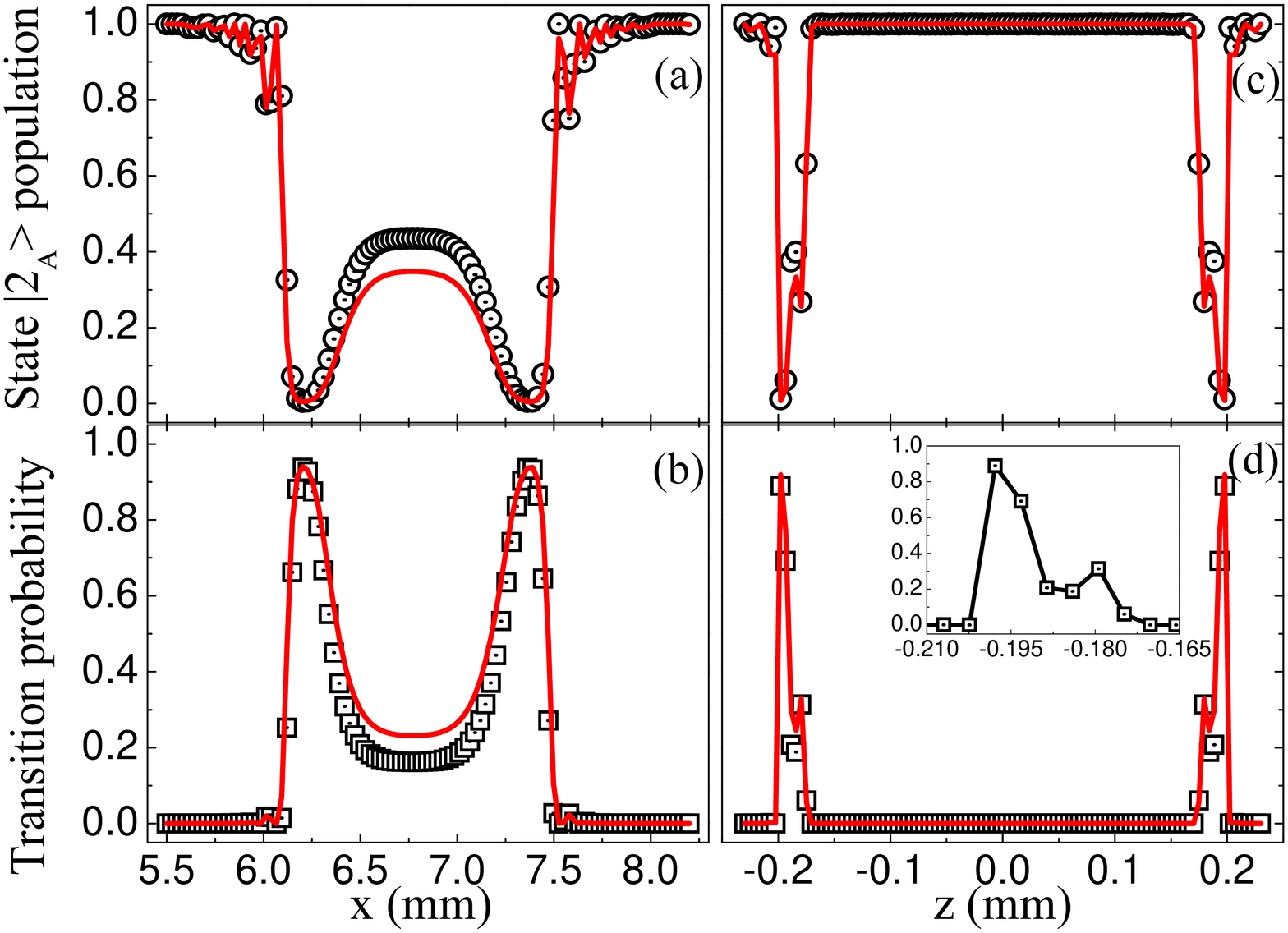}
\caption{(color online). Population of the dressed state $|2_A\ra$
  versus position in the $x$ (a) and $z$ (c) directions at $\Omega_1 t = 20\pi$. The probability for
  an atom to be in the untrapped dressed states is shown in (b) for
  the $x$ and in (d) for the $z$ spatial dependence (solid line --
  Magnus solution, circles and squares -- numerical solution). As before,
  $\Delta = 1.05\Omega_1$ and $\Omega_2 = 0.05\Omega_1$. The inset
  shows a zoom of the resonances around the avoided level crossing at
  $z_{min} \approx -0.19$ mm. The solid line in the inset is just to guide the
  eyes.}
\label{fig:5}
\end{center}
\end{figure}

Away from the resonant conditions just described the analysis of
Eqs.~(\ref{eq:magnusproj1})--(\ref{eq:magnusproj3}) is more
complicated and consequently we evaluate these equations
numerically. Some results for the time evolution of the populations of the adiabatic
states $|0_A\ra$ and $|2_A\ra$ are given in Fig.~\ref{fig:4}. We show
both Magnus approximation (dashed lines) and the exact numerical
solution (circles) for comparison. As can be seen, if we constrain the dressed
spin dynamics to half of the first period, we obtain a very good
agreement between the first order Magnus series and the
exact numerical results. After this time we see dephasing
between the two evolutions and, even more dramatic, an important
disagreement in the amplitude of the observed oscillations. These two
behaviours are somehow expected because at time instants very far from $t
= 0$ the contribution of the next order terms in the Magnus series
becomes more relevant~\cite{milfeld}. In Sec.~\ref{sec:2RWA} we can
find a better approximation (using a second RWA) to the numerical
solution which is shown in Fig.~\ref{fig:4} with solid lines.

Another test for the validity range of the first order Magnus series
approximation is presented in Fig.~\ref{fig:5}, where the population
of the dressed state $|2_A\ra$ and the probability of leaving the rf
trap are shown. In this figure the results of the numerical treatment and
those from the Magnus series are represented by the points and the
solid lines, respectively. To obtain the results in Fig.~\ref{fig:5},
the initial adiabatic spin state has evolved during a time interval approximately equal to one half
of the first oscillation period observed in Fig.~\ref{fig:4}, in short, up
to $\Omega_1 t = 20\pi$ that is $t = \pi/\Omega_2$. Besides the good agreement that both
methods show in the regions of less interest for us, we can notice the
four resonances in the $z$ dependence. The inner peaks are rather
smaller than the outer ones, indicating that we have a
position--dependent resonant coupling.

\subsection{Effective Hamiltonian from a second rotating wave approximation}\label{sec:2RWA}
The first order Magnus approximation predicts well the location of the
resonances and their spin rotation frequency. However, it fails to
describe correctly the dynamics away from the resonance points $\Delta
= \pm\Omega$. To tackle this problem, we used a different
approach and derived more generally applicable analytical
expressions. The approach is based on a
second rotating wave approximation, performed on Eq.~(\ref{eq:iniH}) and expressed through the
rotation $\overline{H}(t) = \mathcal{R}_\Delta H(t)
\mathcal{R}_\Delta^{\dag}$ where $\mathcal{R}_\Delta = \exp(i \Delta t
F_\theta/\hbar)$. This transformation leads to the time--dependent
Hamiltonian $\overline{H}(t)$ given by
\begin{widetext}
\begin{eqnarray}
 \overline{H}(t) = -\big[\Delta-\Omega-\Omega_2 \sin(\theta) \cos(\Delta
 t) \big] F_\theta + \Omega_2\big[ \cos(\theta) \cos^{2}(\Delta
 t)+\sin^{2}(\Delta t) \big] F_{\perp \theta} + \nonumber \\
 \Omega_2 \big[1-\cos(\theta)\big] \sin(\Delta t) \cos(\Delta t)
 F_Y\ ,
\label{eq:adiaH}
\end{eqnarray}
\end{widetext}
where all the parameters appearing in it have already been
introduced. We note that this equation is valid for
any value of $\Delta$, including those close to $\Omega_1$, i.e., when
$\omega_1$ and $\omega_2$ are not so different from each other. Now we
apply a second rotating wave approximation, which is valid provided
that the ``detuning'' $\Delta-\Omega$ and the maximum coupling
$\Omega_2$ are much less than the ``carrier frequency'' $\Delta$. We
finally obtain from Eq.~(\ref{eq:adiaH}) the effective Hamiltonian
\begin{equation}
 \overline{H}_\Delta = -(\Delta-\Omega) F_\theta + \frac{\Omega_2}{2} \big[1+\cos(\theta)\big] F_{\perp \theta} \ .
\label{eq:taH}
\end{equation}
This last equation provides a new compact and powerful description of
the spin dynamics in the dressed trap in the presence of a second rf
field. As an example we have shown in Fig.~\ref{fig:4} the spin
evolution (solid lines) predicted with the effective Hamiltonian which is in
excellent agreement with the exact numerical calculations.

\begin{figure}[htb]
\begin{center}
\includegraphics*[width=0.4\textwidth]{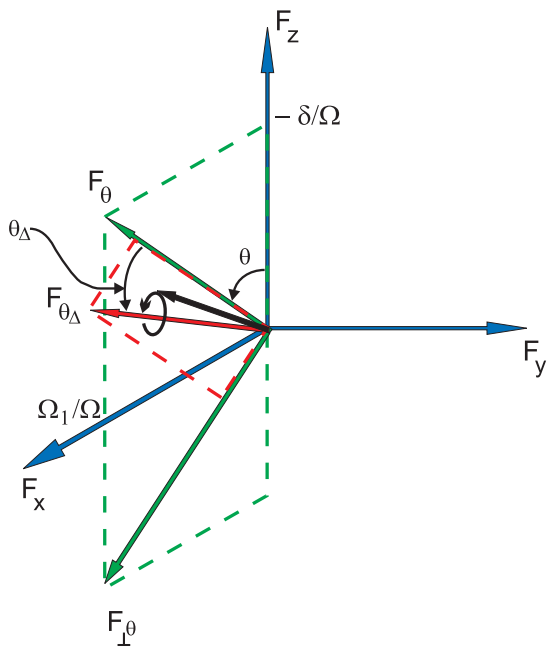}
\caption{(color online). In the presence of the second rf field and
  applying a second RWA, the spin (black arrow) precesses
 around an axis obtained when: Firstly, $\omega_1$ tilts $F_Z$ by $\theta$
 getting $F_\theta$. Secondly, $\omega_2$ tilts $F_\theta$ by
 $\theta_\Delta$ getting $F_{\theta \Delta}$.}
\label{fig:6}
\end{center}
\end{figure}

The form of the effective Hamiltonian (\ref{eq:taH}) is completely equivalent to that of $H_A$ in
Eq.~(\ref{eq:HAapp}). If we look, for instance,
at the vectorial representation of the spin in the case of a single rf
field (Fig.~\ref{fig:1}), then in the presence of the second rf field one
gets the picture shown in Fig.~\ref{fig:6}, where now $F_{\theta
  \Delta} = \cos(\theta_\Delta) F_{\theta} +
\sin(\theta_\Delta) F_{\perp \theta}$ plays a similar role as
$F_\theta$ did before. The angle $\theta_\Delta$ is then
\begin{equation}
 \tan(\theta_\Delta) \equiv -\frac{\Omega_2
   \big[1+\cos(\theta)\big]}{2(\Delta-\Omega)} \ ,\ \mbox{with}\
 \theta_\Delta\in[0,\pi]\ .
\label{eq:thetaD}
\end{equation}
We can view the resulting precession as a second dressing of the
dressed spin states~\cite{ficek}. One can obtain the
doubly--dressed states by diagonalizing the Hamiltonian (\ref{eq:taH}). The
corresponding eigenenergies of these states are given by multiples of $\hbar
\Omega_\Delta$ where clearly
\begin{equation}
 \Omega_\Delta = \sqrt{(\Delta-\Omega)^2 +
   \frac{\Omega_2^{2}}{4}\big[ 1+\cos(\theta) \big]^{2}}\ .
\label{eq:dW}
\end{equation}
The spin oscillation frequency observed in Fig.~\ref{fig:4} is
precisely $\Omega_\Delta$. On resonance, the
period of these Rabi oscillations induced by the coupling in
(\ref{eq:taH}) is then $T = 4\pi/\Omega_2\big[1+\cos(\theta)\big]$ in
agreement with the prediction of the Magnus approximation.

Looking at Fig.~\ref{fig:6} we realize that $\theta_\Delta = \pi/2$
corresponds to resonant coupling with a maximum transition probability to a state orthogonal to
the eigenstates of $F_\theta$. This happens when $\Omega = \Delta$ [see
Eq.~(\ref{eq:thetaD})]. As in the Magnus case, this condition is
realized by the two values $\theta_0 = \arcsin(\Omega_1/\Delta)$ and $\pi-\theta_0$
with $\theta_0\in[0,\pi/2]$. Recalling that when $\theta = \pi/2$
we are exactly at the avoided level crossings in Fig.~\ref{fig:2}(b),
$\theta_0$ indicates the location of the outer resonances OR while $\pi -
\theta_0$ takes care of the inner ones IR.

\section{Evaporation}\label{sec:evap}
\subsection{General remarks}\label{sec:genevap}
In the following two subsections we will consider two schemes for
implementing the evaporative cooling. Firstly, we will look at a pulsed scheme
in which a fraction of the atoms are spin flipped out of the trapped
state by a sudden switch off of the second rf field. Then, secondly,
we will examine a continuous scheme in which hot atoms are removed
from the rf trap by adiabatically following a doubly--dressed state. In
both these schemes the hot atoms that are going to be removed have to
reach the resonances at $\theta_0$ or $\pi-\theta_0$. If we can
neglect the gravitational potential, the energy of
these atoms should be larger than about $F\hbar(\Delta-\Omega_1)$ with respect to
the bottom of the dressed rf trap. This approximation, valid for
relatively small $\Omega_2$, can be refined using
Eq.~(\ref{eq:dW}) and taking gravity into account. As an example, for
our typical experimental setup~\cite{colombe}
and $\Omega_2 = 0.05\Omega_1$, this energy is equivalent to
temperatures of 0.21, 6.1 and 11 $\mu$K for $\Delta$ equal to
1.05$\Omega_1$, 1.25$\Omega_1$ and 1.4$\Omega_1$, respectively.

\begin{figure}[htb]
\begin{center}
\includegraphics*[angle=-90,width=0.45\textwidth]{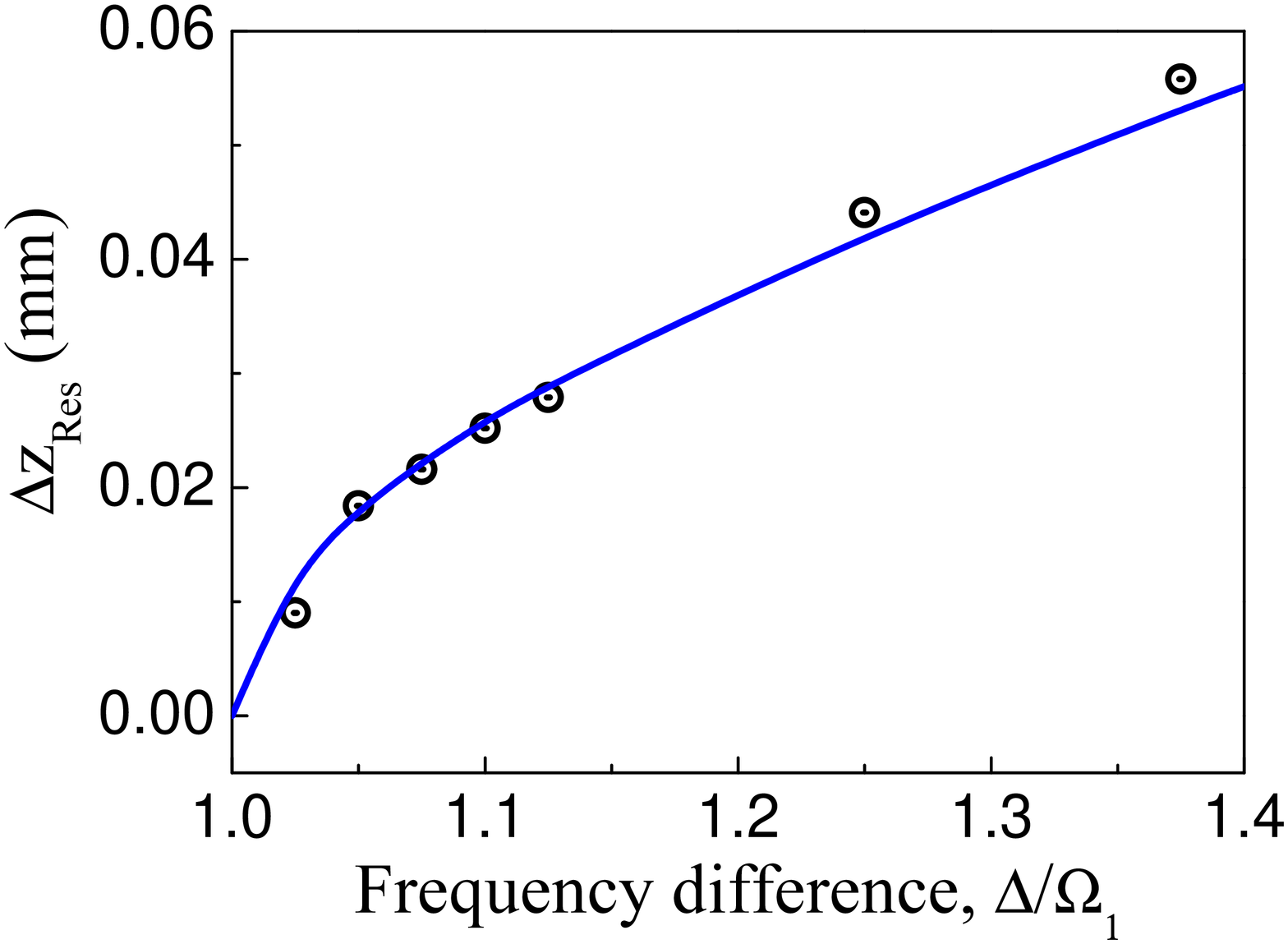}
\caption{(color online). Distance between neighbouring inner and outer
  resonances around an avoided level crossing. The circles are from
  the numerical calculation and the solid line is obtained using
  Eq.~(\ref{eq:dZres}) with $b^{\prime} = 203$ Gauss/cm. The points
  were obtained at $t = \pi/\Omega_2$ by measuring the distance
  between the resonances as seen in the inset of Fig.~\ref{fig:5}(d).}
\label{fig:7}
\end{center}
\end{figure}

These limiting energies imply, of course, that the atom cloud will
have a finite size determined by the location of the inner and
outer resonances. In Fig.~\ref{fig:7} we investigate the distance $\Delta z_{Res}$ between
the neighbouring inner and outer resonances as a function of
$\Delta$. As expected, the distance between the resonances goes to
zero when $\Delta$ is reduced. In fact, since $\Omega_2 \ll \Delta$,
and if we assume a constant magnetic field gradient $b^{\prime}$, we can derive the
approximate form of $\Delta z_{Res}$ from $\Delta=\Omega$ as
\begin{equation}
 \Delta z_{Res} = 2\frac{\sqrt{\Delta^2 - \Omega_1^2}}{\alpha} \ ,
\label{eq:dZres}
\end{equation}
where $\alpha = g_F \mu_B b^{\prime}/\hbar$. 

The last point we would like to consider here concerns the inner
resonance observed in the $z$ direction. One positive aspect about
this resonance is that when $\Delta$ is such that both the inner and
outer resonances are close to the bottom of the dressed rf trap, the
atomic cloud trapped in the adiabatic state $|2_A\ra$ can be
evaporated from both sides. However, the
negative point is that some atoms are transferred by the inner
resonance into the state $|-2_A\ra$ and trapped around $z = 0$. If
these atoms come back to the region of the avoided level
crossing, then they will be energetic enough that heating of the
coldest atoms will take place via inter--atomic collisions. Note that
for the rf dressed trap geometry discussed here gravity favours the
evaporation through the outer resonance because a lower
atomic energy is required than for the inner resonance.

\subsection{Pulsed evaporative cooling}\label{sec:pulseevap}
We first look at the pulsed evaporation scheme, which has as its main
idea the extraction of hot atoms, in a controlled way, from the rf trap
at the resonance locations. We can do this because at these locations
we have large transition probabilities between the rf dressed states as seen in
Figs.~\ref{fig:3}(e), (f). These transitions have been already
analyzed, firstly, by using the Magnus approximation
(Sec.~\ref{sec:Magnus}) and, secondly, by using the second
RWA (Sec.~\ref{sec:2RWA}). They have also been interpreted with the vector
picture in Fig.~\ref{fig:6} as rotations about $F_{\theta
  \Delta}$. Hence we can see that if at a
given time instant $t$ the precession axis of $|\Psi(t)\ra$ has zero
projection onto $F_\theta$, then this state vector will be orthogonal to the
initial dressed rf state and consequently, a transition has taken
place. Clearly, the second rf field can transfer hot atoms out of the
initial trapped dressed state. The pulse has to be repeated several
times during the trap oscillation period to ensure an efficient
evaporation of all the atoms with sufficient energy to reach the
resonances. The pulsed evaporative cooling scheme requires that we can
discriminate between hot and cold atoms by affecting as little as
possible the atoms in the region of the rf
trap centre. This implies a careful choice of the pulse
duration and amplitude as will be discussed below.

\begin{figure}[htb]
\begin{center}
\includegraphics*[angle=-90,width=0.45\textwidth]{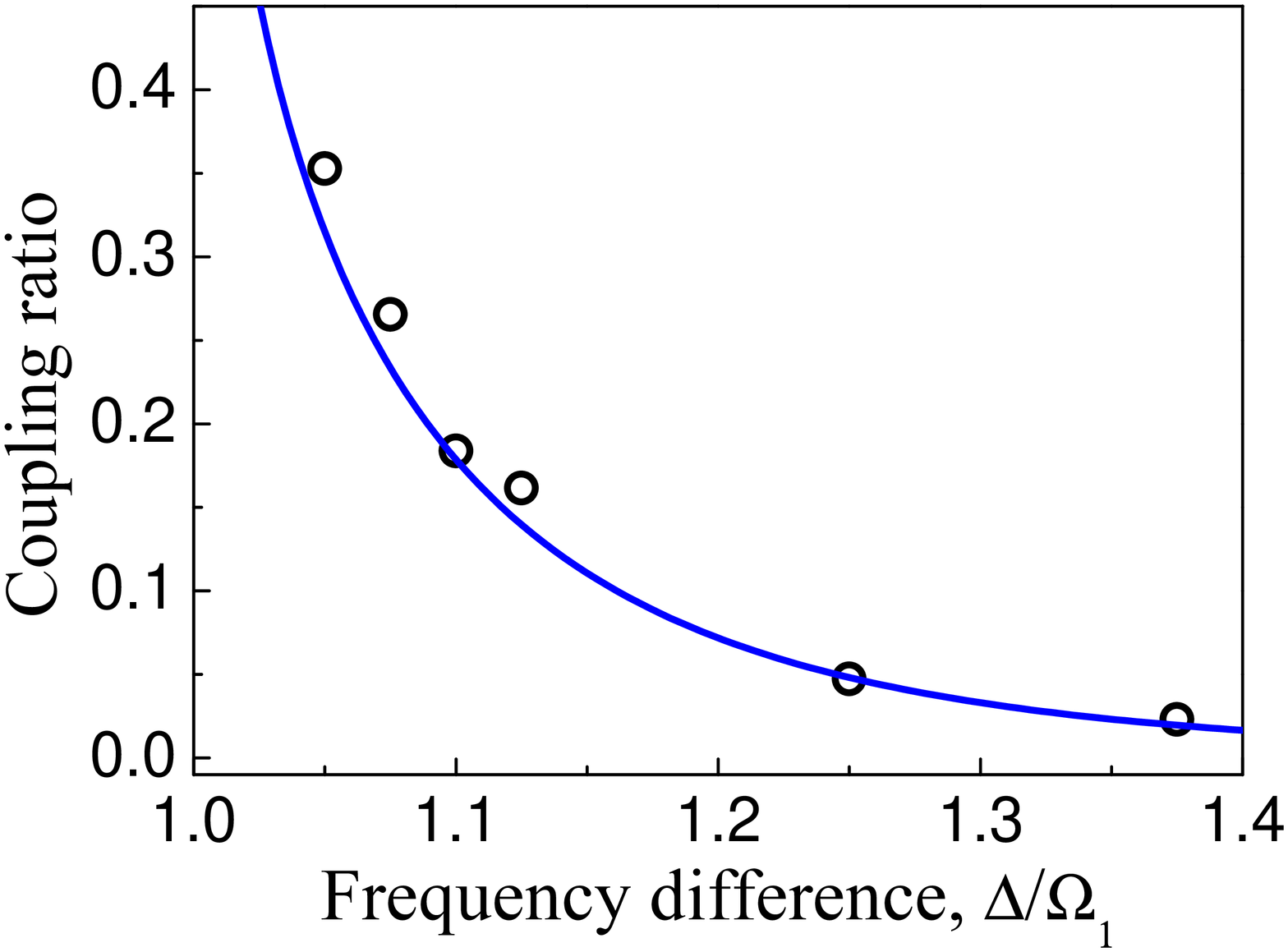}
\caption{(color online). Ratio between the coupling strengths at the
  locations of the inner and outer resonances. The exact numerical
  solution is represented by the points whereas the solid line is
  derived from the second RWA with $\Delta = \Omega$. For $\Delta$ large
  compared to $\Omega_1$, the coupling at the inner resonances goes to
  zero.}
\label{fig:8}
\end{center}
\end{figure}

We already noted that in some situations it may not be desirable to
evaporate via the inner resonance (Sec.~\ref{sec:genevap}). One way we
can avoid this resonance in the pulsed scheme is to carefully chose
the time duration of the pulse. For example, we can see in
Fig.~\ref{fig:3}(f) that the depletion at the inner resonance (IR) takes place
later compared to the outer resonance (OR). This happens because,
independently of the orientation of the dc magnetic field, the
coupling with the Rabi frequency $\Omega_2$ in (\ref{eq:taH}) is
spatially inhomogeneous since $\theta$ depends on $\tbf{r}$. To
investigate this inhomogeneity for different values of $\Delta$, we have
chosen in Fig.~\ref{fig:8} a particular value $t = \pi/\Omega_2$
and explore the ratio of the probability of being in the untrapped
states at the two resonances. This is shown in the figure as a
function of the frequency difference $\Delta$. The second RWA
can predict this relative effectiveness of the inner and outer
resonances. From Eq.~(\ref{eq:taH}) this ratio can be determined
analytically for $\Delta = \Omega$ and it is plotted
in Fig.~\ref{fig:8} with a solid line. The calculation of the numerical results in
Fig.~\ref{fig:8} is done as follows. For a given value of $\Delta$ a figure similar to
Fig.~\ref{fig:5}(d) is plotted. Then, the ratio of the inner peak
height to the outer peak height is found. This is the coupling
strength ratio we are interested in. Since there is an excellent agreement between the
exact solution and the solid line in Fig.~\ref{fig:8}, we can state
that indeed the second RWA works well in the
parameter range we have explored. Notice that for $\Delta \gg
\Omega_1$ we recover the expected result that only the resonant
coupling at the frequency $\omega_2$ will occur between the bare
states. For intermediate values of $\Delta$, we can clearly see that
the particular choice $t = \pi/\Omega_2$ for the pulse duration allows
a good discrimination between the resonances IR and OR. In the final
evaporation stage, where $\Delta \gtrsim \Omega_1$, if we wish to limit
the IR excitation it is necessary to adapt the pulse duration.

\begin{figure}[htb]
\begin{center}
\includegraphics*[angle=-90,width=0.45\textwidth]{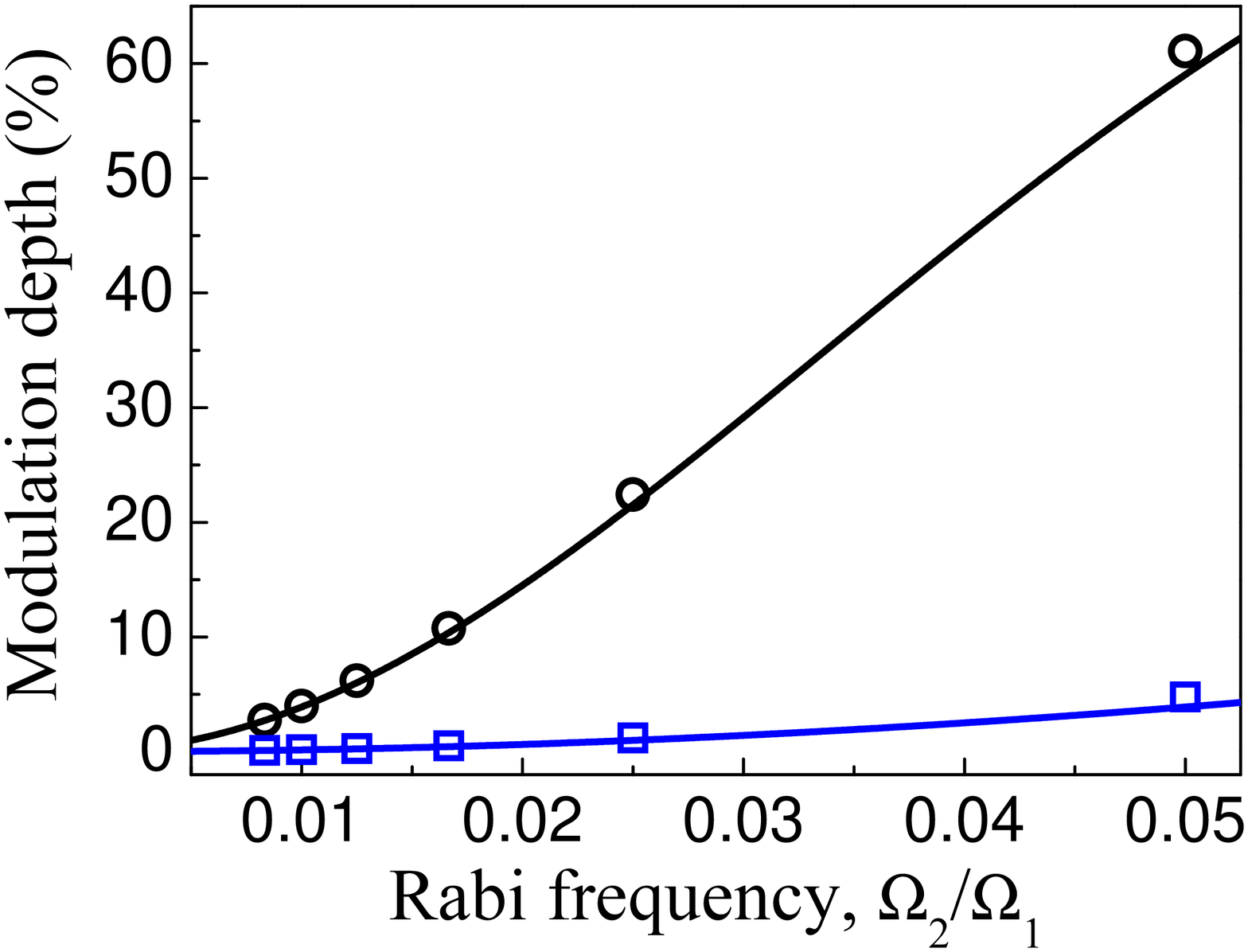}
\caption{(color online). Modulation of the dressed state $|2_A\ra$
  population at the trap centre for $\Delta = 1.05\Omega_1$ (black open
  circles) and 1.25$\Omega_1$ (blue open squares). The solid lines are
  calculated analytically from the second RWA treatment.}
\label{fig:9}
\end{center}
\end{figure}

As mentioned above we should avoid introducing transitions at
the dressed rf trap centre. Because of such transitions we can see in
Figs.~\ref{fig:3}(a), (b), (e) and (f) that, even when the detuning for
coupling adiabatic states is red or blue ($\Delta < \Omega_1$ or $\Delta > \Omega_1$),
the population of the initial trapped dressed state $|2_A\ra$ is in
fact modulated at the centre of the rf trap (avoided level crossing). Such
a modulation can produce unwanted heating or losses and, to
study this process, we introduce the
modulation depth. This quantity is defined as the contrast of the oscillations
presented in Fig.~\ref{fig:4}(a) as the black points (numerical
result). In Fig.~\ref{fig:9} we plot the modulation depth as a function
of the Rabi frequency of the evaporation rf field, and for two
different values of $\Delta$. As expected, the modulation depth
increases with $\Omega_2$ since larger rf power is available for
coupling the adiabatic states. Also not very surprising, this modulation is
more pronounced when $\Delta$ is such that $\omega_2$ couples
adiabatic states close to the position of the avoided level crossing. 

\begin{figure}[htb]
\begin{center}
\includegraphics*[angle=-90,width=0.45\textwidth]{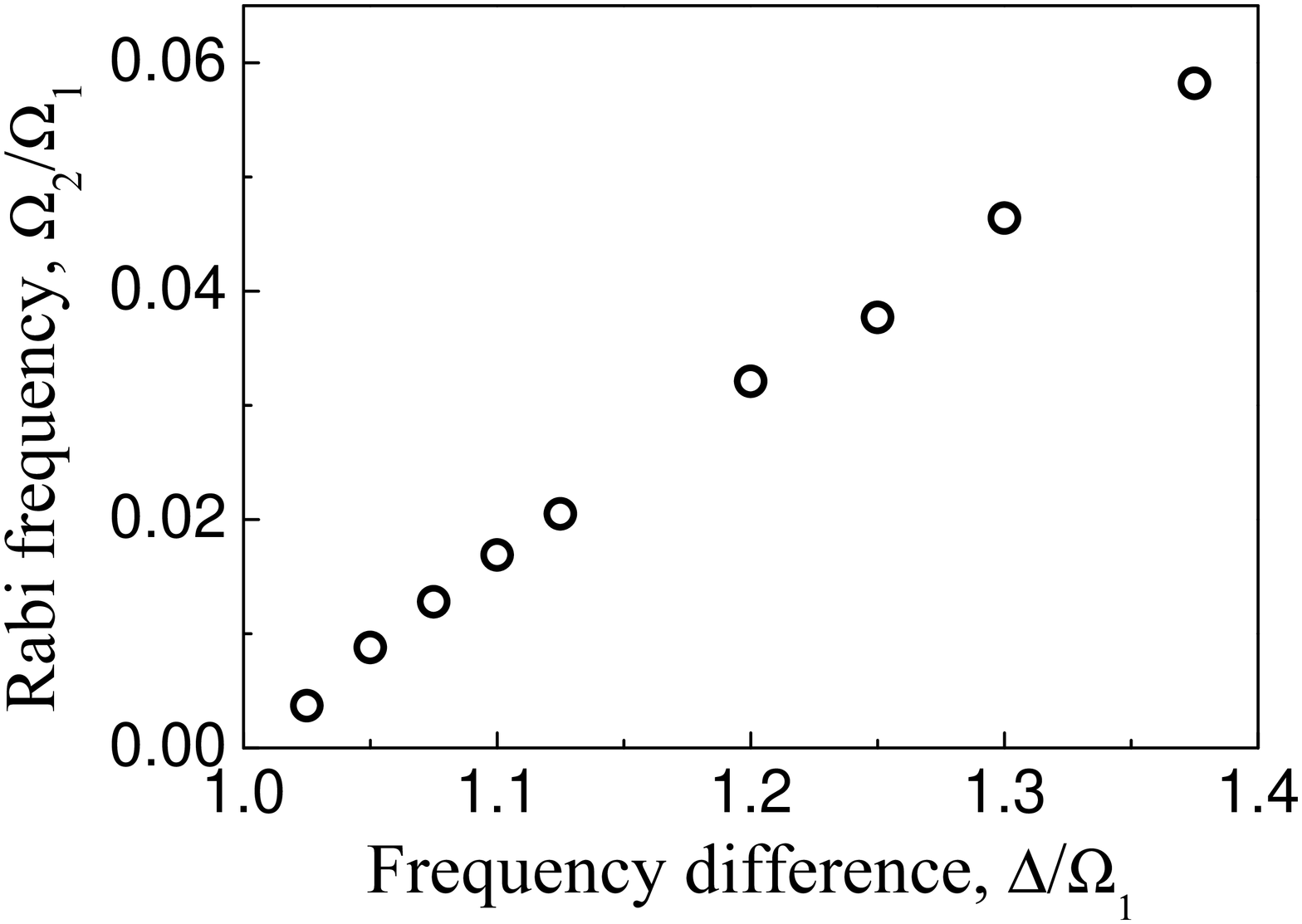}
\caption{Optimized Rabi frequency as a function of the frequency
  difference $\Delta$. In this optimization, the population modulation at the rf
  trap centre is limited to 3\%.}
\label{fig:10}
\end{center}
\end{figure}

Taking into account the result presented in Fig.~\ref{fig:9}, we devised a
strategy which affects as little as possible the coldest
atoms, while doing the evaporation. The idea is to ramp $\Omega_2$ and
$\Delta$ \emph{simultaneously} to preserve a fixed modulation depth at
the rf trap centre~\cite{strat}. In Fig.~\ref{fig:10} such a ramp is presented,
where we have allowed for a 3\% modulation level of the coldest atoms
population. Note that the reduction of $\Omega_2$
has to be taken into account for the optimization of the pulse
duration.

In fact, since $\Omega_2$ and $\Delta$ can be controlled
independently, we can make individual ramps for each one of these
parameters and manipulate independently the position and strength of
the evaporation resonances. Taking into account that there is a finite
modulation of the atomic population at the rf trap centre, even
for $\Delta$ very different from $\Omega_1$, we propose to use a limited
number of rf pulses to cool down the sample. For example, a pulsed
evaporative cooling scheme has been developed demonstrating the achievement of the
collisional regime in a beam of neutral atoms~\cite{lahaye}. However, this
pulsed scheme uses pulses longer than the trap oscillation period
unlike the scheme proposed here.

\subsection{Continuous evaporative cooling}\label{sec:contevap}
Our second scheme for the evaporative cooling of atoms in the
dressed rf trap uses the second rf field in a continuous rather
than pulsed mode. This situation is closer to the normal case of
the evaporation of atoms in a dc magnetic trap by a single rf field.
Here, we simply use the \emph{second} rf field as a tool to
control the dressed rf trap depth. This trap depth corresponds to the
energy required to reach the resonances (with respect to the
dressed rf trap bottom) as considered in Sec.~\ref{sec:genevap}.

\begin{figure}[htb]
\includegraphics[angle=-90,width=0.45\textwidth]{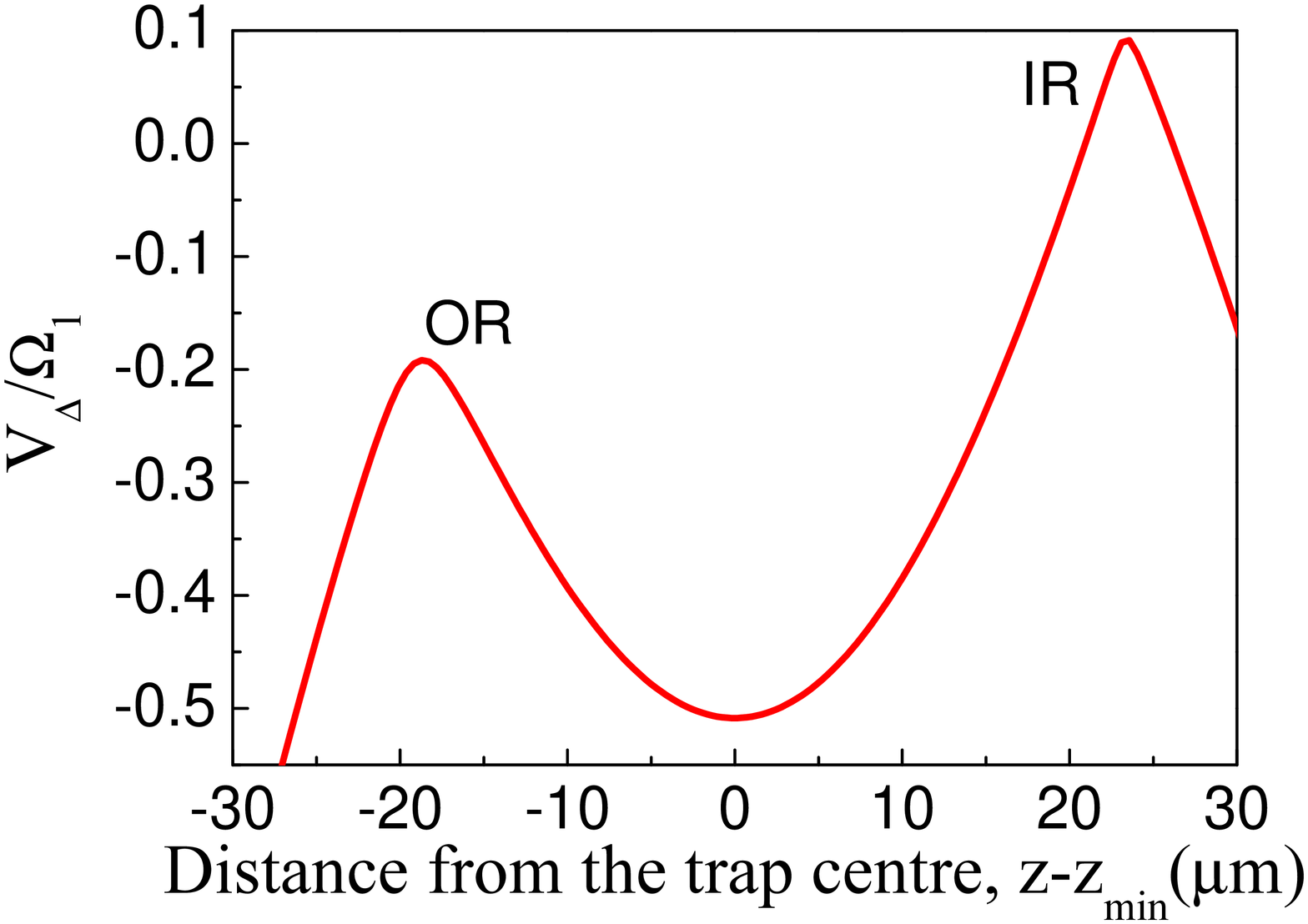}
\caption{(color online). Doubly--dressed potential $V_{\Delta}$
  shown as a function of distance $z-z_{min}$ from the dressed rf trap
  centre at $z_{min}$ for  $\Omega_2=0.05\Omega_1$ and $\Delta=1.25\Omega_1$ as in
  Figs.~\ref{fig:3}(e) and (f). The potential is only
  shown in the region of the singly dressed rf resonance where
  $\delta=0$, with the associated inner and outer doubly--dressed
  resonances indicated with IR and OR as in Fig.~\ref{fig:3}(f). Only
  the relevant, i.e.\ lowest, state is shown for $F=2$ and gravity was
  taken into account in this figure. The
  vertical unit corresponds to about 20 $\mu$K for the value
  of $\Omega_1$ given in Sec.~\ref{sec:conf}. }
\label{fig:11}
\end{figure}

In the usual case of continuous rf evaporation it is
useful to look at the system using dressed states. In our situation,
the equivalent relevant basis is given by doubly--dressed states. As remarked earlier,
these are found by diagonalizing $\overline{H}_\Delta$,
Eq.~(\ref{eq:taH}), or we can write
\begin{equation}
  \overline{H}_\Delta =  \Omega_\Delta F_{\theta \Delta}\ ,
\label{eq:taH1}
\end{equation}
with the frequency $\Omega_\Delta$ as given in Eq.~(\ref{eq:dW}).
This frequency determines the doubly--dressed potential $\hbar
\Omega_\Delta(\mathbf{r})$ which we would like the atoms to follow for the
evaporation to proceed. If we include gravity the resulting
potential $V_{\Delta}(\tbf{r})$ is given by
\begin{equation}
 V_{\Delta}(\tbf{r}) = -F\hbar\Omega_\Delta(\mathbf{r}) + M g z\ .
\label{eq:ddp}
\end{equation}
An example of this potential is given in
Fig.~\ref{fig:11} which shows the minimum at $z = z_{min}$ (i.e.\ at
$\theta=\pi/2$ or $\delta=0$), where the cold atoms will eventually collect, as
well as the resonance regions IR and OR which form the ``lips''
of the doubly--dressed trap over which the hotter atoms must pass.
(During the usual evaporation process the ``lips'' are
subsequently lowered by ramping the rf frequency and the same
procedure can be carried out here with the second rf field.)

For this picture to be valid, we must have an adiabatic following
of the vector $F_{\theta \Delta}$ as the atoms move about the
trap. A general condition for this can be expressed as
\begin{equation}
 \Big|\frac{d\theta_\Delta}{dt}\Big| \ll \Omega_\Delta \ ,
\label{eq:adiacond}
\end{equation}
where $\theta_\Delta$ is given in Eq.~(\ref{eq:thetaD}) and the
right--hand member of the inequality (\ref{eq:adiacond}) is seen
to be just the energy separation between the doubly--dressed
levels. In practice this condition is rather easily satisfied for
a singly dressed rf trap (see, for example, Ref. \cite{colombe}), which
is relevant for the bottom of the doubly--dressed trap as
illustrated in Fig.~\ref{fig:11}. However, in order to also
satisfy the adiabatic following condition (\ref{eq:adiacond}) at the
resonances, we will find a new constraint
that $\Omega_2$ should not be \emph{too} small. The analysis of
adiabaticity is conveniently carried out in terms
of a Landau-Zener parameter $\Lambda$ such that Eq.~(\ref{eq:adiacond}) implies
that $\Lambda \gg 1$. To proceed, we use the definition
(\ref{eq:thetaD}) of $\theta_\Delta$ in order to compute its time
derivative assuming that only $\theta$ is time--dependent.
Calculating the time derivative and evaluating
(\ref{eq:adiacond}) at the outer resonance location $\theta_0$, we find
that in terms of the Landau--Zener parameter $\Lambda =
\Omega_\Delta/|\dot{\theta}_\Delta|$ the
adiabatic following condition reads
\begin{equation}
 \Lambda = \frac{\Omega_2^{2}\big[ 1+\cos(\theta_0)
   \big]^{2}}{4|\dot{\delta}_0|\cos(\theta_0)} \gg 1 \ ,
\label{eq:LZ}
\end{equation}
where $\dot{\delta}_0$ is the time derivative of the detuning
$\delta(\tbf{r})$ evaluated at $\theta_0$, and is proportional to
an atom velocity. From the expression for $\Lambda$ we can see
that as soon as $\Omega_2$ is reduced, only the slow atoms will
have their spin adiabatically following the axis $F_{\theta
  \Delta}$, and in fact, for motion linearized over the
resonance, the multi-state Landau-Zener analysis \cite{vitanov}
shows that the probability for an atom to be lost from the
adiabatic state in a single pass is
$ 1 - [ 1 - \exp(-\pi\Lambda)]^{2F}$. For the parameters of Fig.~\ref{fig:11},
the energy of the atoms $E$ (measured from the trap bottom in
temperature units) would have to have
a value of $E/k_B = 20$  $\mu$K for the probability of a non--adiabatic
crossing to reach about $10^{-5}$ for the outer resonance OR.

It is clear that if we switch from the outer resonance OR (at
$\theta=\theta_0$) to the inner resonance IR (at
$\theta=\pi-\theta_0$) the adiabaticity condition (\ref{eq:LZ})
will be different. This is connected to both the spatially
dependent coupling $\Omega_2 \big[1+\cos(\theta)\big]/2$, and the
effect of gravity, which as seen in Fig.~\ref{fig:11}
make the IR ``lip'' higher. The weaker coupling at the inner
resonance also means that the dynamics is less adiabatic at this
point. In fact for fairly ``hot'' atoms one can contrive that the
resonance OR is rather adiabatic whilst the resonance IR is
rather diabatic. Together with the effect of gravity, this would mean
that atoms can be evaporated out
of the OR resonance whilst adiabatic coupling through the IR
resonance is prevented. As explained in
Sec.~\ref{sec:genevap} this can be useful to partially prevent the return of
evaporated atoms to the resonance regions with subsequent
collisions and heating. However, we note that if we want to
reduce the final temperature by steadily reducing $\Delta$, the
couplings at the two resonances become more equal (as in the
pulsed scheme) and less discrimination between the two
evaporation zones is possible.

\section{Conclusions}\label{sec:conclu}
We have seen that we can employ a doubly--dressed basis~\cite{ficek}
for the analysis of a dressed rf trap with two rf fields
provided the second rf field is sufficiently weak. Using this
arrangement of fields we can create a scheme for the
evaporative cooling of atoms in a singly dressed rf trap in a
continuous mode. In contrast with the traditional continuous forced evaporation scheme
the idea of evaporative cooling based on the application of rf pulses
with properly chosen durations and frequencies is also developed for a
dressed rf trap. The duration of such
pulses is essentially determined by the power of the rf field used for the
evaporation, although the optimal pulse length changes from one
location to another in the adiabatic trapping potential. When the
trapping and evaporative cooling radio--frequencies are comparable, we
found the evaporation to happen \emph{also} via additional
resonances. Even if these resonances can enhance the evaporation process in both the pulsed
and continuous schemes, we have to be careful that atoms
evaporated through these resonances do not come back and heat the cold atomic
cloud. In this respect, the advantage of the pulsed scheme is to
provide an additional control of the transition probability at the
inner resonance via the pulse
duration. However, the effect of gravity is to favour the evaporation via
the outer resonance OR which is then located at a position of lower
energy. Although the importance of this effect may depend on the
particular trapping geometry (pancake, ring traps), the evaporation
schemes proposed in this paper are quite general and should allow an
efficient cooling of atoms
directly in the various rf traps which have been proposed or 
realized~\cite{colombe,schumm,morizot,courteille,lesanovsky,fernholz}.

The main application of the work in this paper is to the
evaporative cooling of atoms in a dressed trap. However, there are also
applications concerning noise and stability of dressed rf traps where 
the carrier frequency $\omega_1$ is not perfectly monochromatic. This
can be due, for instance, to contamination by stray fields. Then it is
clear that if the frequency components next
to the carrier are in the range of, let's say,  $\Delta =
1.01\Omega_1$ -- $1.05\Omega_1$, they will empty the rf trap if they
have enough power to do so or, in the best case,
they will raise the temperature of the atoms. This means that if we look at the second
rf source as a noise term (a sideband in the frequency spectrum), the
results presented here can be used to estimate the damage it causes.

\begin{acknowledgments}
We acknowledge the financial support of the R\'egion Ile-de-France
(contract No. E1213). CLGA acknowledges the support from a Marie Curie
fellowship (``Atom Chips'', MRTN-CT-2003-505032). BMG thanks the
University Paris 13. Laboratoire de Physique des Lasers is UMR 7538 of
CNRS and University Paris 13. The LPL group is a member of the
Institut Francilien de Recherche des Atomes Froids (IFRAF).
\end{acknowledgments}

\appendix
\section{Derivation of $H(\tbf{r},t)$}\label{app:Hamiltonian}
We will start the derivation of Eq.~(\ref{eq:iniH}) by considering
that the total magnetic field experienced by the atoms consists of
three contributions or terms. One coming from the inhomogeneous dc
magnetic field $\tbf{B}_{dc}(\tbf{r})$ of the QUIC trap, a second term oscillating at
the frequency $\omega_1$, $\tbf{B}_1(\tbf{r},t) = B_{01} \tbf{e}_1 \cos(\omega_1 t)$, associated with
the adiabatic trapping potential, and a third term of frequency $\omega_2$,
$\tbf{B}_2(\tbf{r},t) = B_{02} \tbf{e}_2 \cos(\omega_2 t)$, responsible
for the evaporative cooling in the rf trap. Here, $B_{01}$ and
$B_{02}$ are the amplitudes of the fields whereas $\tbf{e}_1$ and
$\tbf{e}_2$ are unit polarization vectors. Using these definitions
and denoting by $\tbf{F}$ the atomic angular momentum operator,
the total Hamiltonian of our physical system $H_T(\tbf{r},t)$ can be
approximated by 
\begin{equation}
 H_T(\tbf{r},t) = g_F \mu_B \tbf{F}\cdot\big[\tbf{B}_{dc}(\tbf{r}) +
 \tbf{B}_1(\tbf{r},t) + \tbf{B}_2(\tbf{r},t)\big]/\hbar \ ,
\label{eq:HT}
\end{equation}
where $g_F$ and $\mu_B$ are the Land\'e factor and the Bohr magneton,
respectively. If we assume that at every point $\tbf{r}$ the direction
of the dc magnetic field defines the local $Z$ quantization axis then,
for $X$ polarized rf fields, the Eq.~(\ref{eq:HT}) takes the form
\begin{equation}
 H_T(\tbf{r},t) = \omega_0(\tbf{r}) F_Z + V_1(\tbf{r},t) + V_2(\tbf{r},t) \ .
\label{eq:HT1}
\end{equation}
In (\ref{eq:HT1}) $\omega_0(\tbf{r}) = g_F \mu_B
B_{dc}(\tbf{r})/\hbar$ is the Larmor precession frequency. The interaction
Hamiltonian $V_j(\tbf{r},t)$ is defined by the expression
\begin{equation}
 V_j(\tbf{r},t) = \Omega_j(\tbf{r}) F_X \big(e^{i \omega_j t} + e^{-i \omega_j t}\big)\ \ \ ,\ \ \ j = 1,2
\label{eq:interV}
\end{equation}
being $\Omega_j(\tbf{r}) \equiv g_F \mu_B B_{0j}/(2\hbar)$ the Rabi
frequency. 

Given $H_T(\tbf{r},t)$, the dynamics of an atomic spin state
$|\phi(\tbf{r},t)\ra$ is governed by the Schr\"odinger equation
\begin{equation}
i \hbar \frac{d|\phi(\tbf{r},t)\ra}{dt} = H_T(\tbf{r},t) |\phi(\tbf{r},t)\ra\ ,
\label{eq:SE}
\end{equation}
which in the frame rotating at the frequency $\omega_1$ becomes
\begin{eqnarray}
 i \hbar \frac{d|\psi(\tbf{r},t)\ra}{dt} = \big[-\delta(\tbf{r}) F_Z +
 \nonumber\\
 \mathcal{R}^{\dag}_1 V_1(t) \mathcal{R}_1 + \mathcal{R}^{\dag}_1 V_2(t) \mathcal{R}_1 \big]|\psi(\tbf{r},t)\ra\ .
\label{eq:SE1}
\end{eqnarray}

In Eq.~(\ref{eq:SE1}) we have introduced the detuning \mbox{$\delta(\tbf{r})
= \omega_1 - \omega_0(\tbf{r})$}, the rotating frame operator
\mbox{$\mathcal{R}_1 = \exp(-i \omega_1 t F_Z/\hbar)$}, and the rotated state
\mbox{$|\psi(\tbf{r},t)\ra = \mathcal{R}^{\dag}_1 |\phi(\tbf{r},t)\ra$}. If we
consider the bare state basis $\{|-2\ra, |-1\ra, |0\ra, |+2\ra,
|+1\ra\}$ of a spin--2 system, the matrix form of the rotated interaction
Hamiltonians $\mathcal{R}^{\dag}_1 V_1(t) \mathcal{R}_1$ and
$\mathcal{R}^{\dag}_1 V_2(t) \mathcal{R}_1$ are respectively given by
\begin{displaymath}
 \Omega_1(\tbf{r}) \left( \begin{array}{ccccc}
 0 & 1 & 0 & 0 & 0 \\
 1 & 0 & \sqrt{\frac{3}{2}} & 0 & 0 \\
 0 & \sqrt{\frac{3}{2}} & 0 & \sqrt{\frac{3}{2}} & 0 \\
 0 & 0 & \sqrt{\frac{3}{2}} & 0 & 1 \\
 0 & 0 & 0 & 1 & 0
 \end{array} \right) 
\end{displaymath}
and
\begin{displaymath}
 \Omega_2(\tbf{r}) \left( \begin{array}{ccccc}
 0 & e^{i \Delta t} & 0 & 0 & 0 \\
 e^{-i \Delta t} & 0 & \sqrt{\frac{3}{2}} e^{i \Delta t} & 0 & 0 \\
 0 & \sqrt{\frac{3}{2}} e^{-i \Delta t} & 0 & \sqrt{\frac{3}{2}} e^{i \Delta t} & 0 \\
 0 & 0 & \sqrt{\frac{3}{2}} e^{-i \Delta t} & 0 & e^{i \Delta t} \\
 0 & 0 & 0 & e^{-i \Delta t} & 0
 \end{array} \right) \ ,
\end{displaymath}
where $\Delta = \omega_2 - \omega_1$ and we have made use of the
rotating wave approximation (RWA) by discarding the terms that
oscillate at $2\omega_1$, $2\omega_2$, and $\omega_1 +
\omega_2$. In general, we
find the dynamics of $|\psi(\tbf{r},t)\ra$ in (\ref{eq:SE1}) to be
described by the Hamiltonian
\begin{equation}
 H(\tbf{r},t) = H_A(\tbf{r}) + \Omega_2 \big[F_X \cos(\Delta t) 
 + F_Y \sin(\Delta t)\big] \ ,
\label{eq:iniHapp}
\end{equation}
where the adiabatic Hamiltonian $H_A(\tbf{r})$ is defined as
\begin{equation}
 H_A(\tbf{r}) = -\delta(\tbf{r}) F_Z + \Omega_1 F_X \ .
\label{eq:HAapp}
\end{equation}

The time--independent Hamiltonian (\ref{eq:HAapp}) can be rewritten as
$H_A(\tbf{r}) = \Omega(\tbf{r}) F_\theta$ if we define $\Omega(\tbf{r}) = \sqrt{\delta(\tbf{r})^{2}+\Omega_1^{2}}$, $\cos(\theta) =
-\delta(\tbf{r})/\Omega(\tbf{r})$, \mbox{$\sin(\theta) = \Omega_1/\Omega(\tbf{r})$}, and $F_\theta = \cos(\theta)F_Z + \sin(\theta) F_X$.



\begin{thebibliography}{99}
\bibitem{strat} Another strategy to reduce the population transfer in the
 rf trap centre could be to include this requirement in the
 optimization of the pulse duration.
\end{thebibliography}
\end{document}